\documentclass[10pt]{amsart}
\usepackage{amssymb}
\usepackage{enumerate}
\usepackage{comment}
\usepackage{epsf}
\usepackage{psfrag}
\usepackage{graphics}
\usepackage[dvips]{graphicx}
\usepackage{xcolor}
\usepackage[colorlinks=true,
            urlcolor=blue]{hyperref}
\usepackage{slashed}
\usepackage{graphicx}
\usepackage{wrapfig}
\usepackage{float}
\DeclareGraphicsExtensions{.eps,.art,.ART,.ps}
\usepackage{mathtools}
\usepackage{graphicx,caption,subcaption}
\usepackage{tikz}
\newcommand{\cL}{\mathcal{L}}

\newcommand{\bbR}{\mathbb{R}}      
\newcommand{\bbZ}{\mathbb{Z}}      
\newcommand{\amplitude}{h}

\newcommand{\tr}{\operatorname{tr}}

\newcommand{\cof}{\operatorname{cof}}
\newcommand{\supp}{\operatorname{supp}}

\newcommand{\Imag}{\operatorname{Im}}

\begin{document}
\begin{abstract}
We study the interaction between gravitational waves and elastic bodies within the framework of relativistic elasticity. Starting from the Lagrangian formulation of relativistic elasticity, we derive the linearized equations governing the response of a homogeneous and isotropic solid to a weak gravitational wave by expanding the action to quadratic order in the elastic field derivatives and to linear order in the metric perturbation. In this way, Dyson’s interaction term from the effective potential approach emerges naturally from the relativistic theory.

We then apply the formalism to a thin rectangular elastic plate aligned with the direction of propagation and polarization of a plus-polarized gravitational wave. For a material with vanishing Poisson ratio, the equations decouple and admit simple explicit solutions. We obtain closed-form expressions for the induced displacements and for the energy deposited on the plate by both short gravitational wave bursts and continuous harmonic waves. Finally, we compute the gravitational wave emission generated by the oscillating plate itself under continuous harmonic excitation. These results provide a fully relativistic derivation of the elastic response to gravitational waves together with explicit solvable examples relevant to resonant gravitational wave detectors.
\end{abstract}
%
%
\title[Relativistic Elastic Response to Gravitational Waves]{Relativistic Elastic Response to Gravitational Waves:
Explicit Solutions for a Rectangular Plate}
\author{Jos\'{e} Nat\'{a}rio and Filipe Nazar\'{e}$^*$}
\thanks{$^*$Corresponding author \href{mailto:filipe.alexandre.nazare.pereira.dos.santos@tecnico.ulisboa.pt}{\tt <filipe.alexandre.nazare.pereira.dos.santos@tecnico.ulisboa.pt>}} 
\address{Center for Mathematical Analysis, Geometry and Dynamical Systems, Mathematics Department, and Center for Astrophysics and Gravitation, Physics Department, Instituto Superior T\'ecnico, Universidade de Lisboa, Portugal}
%
%
\maketitle
%
%
%
%
\thispagestyle{empty}
%
%
%
\section*{Introduction}

The direct detection of gravitational waves by the LIGO and Virgo collaborations \cite{LIGOScientific:2016aoc} marked a turning point in modern physics: not only did it confirm a fundamental prediction of general relativity, first derived by Einstein \cite{Einstein:1916cc}, but, more importantly, it opened a new observational window onto the universe. With the upcoming Laser Interferometer Space Antenna (LISA), gravitational wave measurements are expected to reach unprecedented levels of precision in the $10^{-4} - 1~\rm{Hz}$ range, probing a wider range of sources and physical effects beyond those accessible to current ground-based detectors (which operate primarily in the $10 - 10^4~\rm{Hz}$ band).

Long before these interferometric detections, gravitational waves were sought using resonant detectors, most notably the bar detectors pioneered by Joseph Weber \cite{Weber:1960zz}. These devices consisted of aluminum cylinders designed to resonate around $10^3~{\rm Hz}$ under the influence of a passing gravitational wave. Although Weber’s reported detections in the late 1960s \cite{Weber:1969bz} were not confirmed, and are now generally regarded as spurious, the conceptual framework he introduced played a foundational role in the development of gravitational-wave detection technology. In particular, these early efforts established the basic principles underlying resonant detectors and motivated continued study of the interaction between gravitational waves and elastic bodies, including the experiments described in \cite{Astone:1993, Astone:1997, Cerdonio:1997}. More recently, there has been renewed interest in the elastic response of matter to gravitational waves \cite{Bi:2024mve, Kachelriess:2023awv, Majstorovic:2024jvl, Yan:2024jio, Vicente:2024ksp}, including the revival of Weber’s original proposal to use the Moon as a giant gravitational-wave detector in the $10^{-3} - 10^{-1}~\mathrm{Hz}$ range.

The theoretical description of resonant detectors naturally leads to the  theory of relativistic elasticity, originally formulated by Carter and Quintana \cite{CarterQuintana72} (see also \cite{Maugin1978, KijowskiMagli1992, TahvildarZadeh1998, Beig:2002pk, Karlovini:2002fc, Wernig-Pichler:2006xdj, Brown:2020pav, Beig:2023pka} and references therein). This theory, which is expected to play a central role in the modeling of neutron stars and other compact objects \cite{Chamel:2008ca, Andersson:2018xmu}, offers a natural setting to describe how perturbations of the spacetime metric, such as gravitational waves, induce measurable deformations in elastic media.

The equations governing the interaction of a homogeneous and isotropic solid with a gravitational wave were first derived by Dyson \cite{Dyson:1969zgf}, who introduced an interaction term into the Lagrangian of Newtonian linear elasticity (the so-called ``effective potential approach'').\footnote{This formalism was subsequently shown to be equivalent to introducing a Newtonian tidal force associated with the Riemann curvature tensor of the gravitational wave; see \cite{Majstorovic:2024jvl}.} They were later derived from first principles within the framework of relativistic elasticity by Carter and Quintana \cite{Carter:1977qf}, and have recently been revisited and extended in \cite{Hudelist:2022ixo, Belgacem:2024lzi}.

In this work, we adopt a more elementary and self-contained approach. We expand the relativistic elastic Lagrangian directly, retaining terms up to quadratic order in the field derivatives and linear order in the metric perturbation, with all coefficients expressed in terms of the rest-frame energy density and the two sound speeds. Dyson's interaction term then emerges naturally from this expansion. The main contribution of our work, however, lies in the explicit analysis of the gravitational-wave response of elastic bodies. In particular, for a thin rectangular plate aligned with the propagation and polarization directions of the gravitational wave, and assuming a material with vanishing Poisson ratio, the problem reduces to two decoupled one-dimensional wave equations subject to Neumann boundary conditions. We solve these equations in closed form for an arbitrary gravitational-wave signal, allowing us to estimate the energy deposited in the plate by short bursts and to derive explicit expressions for its response to continuous monochromatic waves. Finally, we calculate the gravitational radiation re-emitted by the plate itself, including configurations in which the leading quadrupole contribution surprisingly vanishes.

The structure of the paper is as follows. After briefly reviewing the necessary concepts in elasticity in Section \ref{section1}, we develop the formalism describing the elastic response to gravitational waves in Section \ref{section2}. In Section \ref{section3}, we apply this theory to a two-dimensional rectangular plate subject to a plus-polarized gravitational wave, obtaining explicit expressions for displacements and energy. Sections \ref{section4} and \ref{section5} analyze the examples of a short gravitational wave burst and a continuous harmonic wave, respectively, with particular emphasis on the energy deposited on the plate. In Section \ref{section6} we study the gravitational wave emission of the plate excited by a continuous harmonic wave. Finally, we summarize our conclusions in Section \ref{section7}.

We mostly follow the conventions of \cite{Misner:1973prb}; in particular, we use a geometrized system of units, in which both the speed of light $c$ and Newton's gravitational constant $G$ are set to unity, $c=G=1$.
%
%
%
\section{Relativistic elasticity}\label{section1}
We start by briefly recalling the basic setup for the Lagrangian formulation of relativistic elasticity (see \cite{Natario:2019nrf} for more details). Given a spacetime $(M,g)$, we model a continuous medium as a Riemannian 3-manifold $(S, k)$ (which we call the {\bf relaxed configuration}) and a projection map $\pi: M \to S$ whose level sets are timelike curves (the worldlines of the medium particles). If we choose local coordinates $(\bar{x}^1, \bar{x}^2, \bar{x}^3)$ in $S$ then we can think of $\pi$ as a set of three scalar fields $\bar{x}^1, \bar{x}^2, \bar{x}^3$ defined on $M$. The deformations of the medium with respect to the natural metric $k_{ij}$ of $S$ are then given by the (inverse) metric
\begin{equation} \label{hij}
h^{ij} = g^{\mu\nu} \partial_\mu \bar{x}^i \partial_\nu \bar{x}^j \, .
\end{equation}

An {\bf elastic} Lagrangian density $\cL$ is one that depends on the derivatives of the fields $\bar{x}^1, \bar{x}^2, \bar{x}^3$ only through $h^{ij}$; the choice of $\cL = \cL(\bar{x}^i, h^{ij})$, which turns out to be the rest energy density measured by a comoving observer, is called the {\bf elastic law} of the continuous medium.

We define {\bf homogeneous and isotropic materials} to be those for which $\cL$ depends only on the eigenvalues of $h_{ij}$ with respect to $k_{ij}$ (that is, the eigenvalues of the matrix $(h_{ij})$ in a frame where $k_{ij}=\delta_{ij}$). If $k_{ij}=\delta_{ij}$, that is, if $(S, k)$ is the Euclidean space, then we can define the more convenient variables
\begin{equation}
\lambda_0 = \det (h^{ij}) \, , \qquad \lambda_1 = \tr (h^{ij}) \, , \qquad \lambda_2 = \tr \cof (h^{ij})
\end{equation}
for the Lagrangian of a homogeneous and isotropic material (where $\cof(h^{ij})$ is the matrix of cofactors of $(h^{ij})$). Note that in the relaxed configuration, that is, when $h^{ij}=\delta^{ij}$, these variables take the values $\lambda_0=1$ and $\lambda_1=\lambda_2=3$.
%
%
\section{Linear elasticity and gravitational waves}\label{section2}
In this section we re-derive the linearized equations of motion for a homogeneous and isotropic solid excited by a gravitational wave. Our approach differs slightly from those in \cite{Carter:1977qf, Hudelist:2022ixo, Belgacem:2024lzi} in that we directly expand the Lagrangian for relativistic elasticity to quadratic order in the field derivatives and to linear order in the metric perturbation. Dyson's interaction term emerges naturally from this expansion, again confirming the validity of the effective potential approach.

We will derive the linearized equations of motion for the metric of a $+$ polarized linear gravitational wave written in the $TT$ gauge,
\begin{equation} \label{metricGW}
g = -dt^2 + \left[ 1 + \varphi(t-z) \right] dx^2 + \left[ 1 - \varphi(t-z) \right] dy^2 + dz^2 \, .
\end{equation}
The equations of motion for a $\times$ polarized wave can be easily obtained by rotating the $xy$ axes by $45^\circ$, and the equations of motion in the general case follow by superposition.

We look for solutions of the equations of motion of the form\footnote{Note that $\xi^i=\bar{x}^i-x^i$ can be thought of as minus the displacement of the particles with respect to their equilibrium positions when $\varphi = 0$, that is, in Minkowski spacetime.}
\begin{equation}
\bar{x}^i = x^i + \xi^i \, ,
\end{equation}
which can be written out in full as 
\begin{equation}
\begin{cases}
\bar{x}(t,x,y,z)=x+\xi(t,x,y,z) \\
\bar{y}(t,x,y,z)=y+\eta(t,x,y,z) \\
\bar{z}(t,x,y,z)=z+\zeta(t,x,y,z)
\end{cases} \, .
\end{equation}
We then have
\begin{equation}
d\bar{x}^i = dx^i + d\xi^i \, ,
\end{equation}
that is,
\begin{equation}
\begin{cases}
d\bar{x}=dx+d\xi \\
d\bar{y}=dy+d\eta \\
d\bar{z}=dz+d\zeta
\end{cases}.
\end{equation}
Let us assume that $|d\xi^i| \ll 1$ (in addition to the standard linearized gravity condition $|\varphi| \ll 1$).
Since the linearized inverse metric is
\begin{equation}
g^{-1} = -\partial_t^2 + \left[ 1 - \varphi(t-z) \right] \partial_x^2 + \left[ 1 + \varphi(t-z) \right] \partial_y^2 + \partial_z^2 \, ,
\end{equation}
we obtain, to second order on the main diagonal and to first order elsewhere,
\begin{align}
\left(\gamma^{ij}\right) & := g^{-1}(d\bar{x}^i, d\bar{x}^j) \\
& = \left(
\begin{matrix}
1 + 2 \xi_x + \left\langle d\xi, d\xi \right\rangle - \varphi (1+2\xi_x) & \eta_x + \xi_y & \zeta_x + \xi_z \\
\eta_x + \xi_y & 1 + 2 \eta_y + \left\langle d\eta, d\eta \right\rangle + \varphi (1+2\eta_y) & \zeta_y + \eta_z \\
\zeta_x + \xi_z & \zeta_y + \eta_z & 1 + 2 \zeta_z + \left\langle d\zeta, d\zeta \right\rangle 
\end{matrix}
\right) \, , \nonumber
\end{align}
where $\langle\cdot,\cdot\rangle$ stands for the Minkowski inner product of covectors. Therefore, to second order in $\xi^i$ and $\varphi$ (but ignoring quadratic terms in $\varphi$), we have
\begin{align}
\lambda_0  = & \,\, \det\left(\gamma^{ij}\right) \nonumber \\
= & \,\, 1 + 2\xi_x + 2\eta_y + 2\zeta_z + 4\xi_x\eta_y + 4\xi_x\zeta_z + 4\eta_y\zeta_z \nonumber \\
& + \left\langle d\xi, d\xi \right\rangle + \left\langle d\eta, d\eta \right\rangle + \left\langle d\zeta, d\zeta \right\rangle \nonumber \\
& - (\eta_x + \xi_y)^2 - (\zeta_x + \xi_z)^2 - ( \zeta_y + \eta_z)^2 \,, \label{eq:lambda0}
\end{align}
as well as
\begin{align}
\lambda_1  = & \,\, \tr\left(\gamma^{ij}\right) \nonumber \\
= & \,\, 3 + 2\xi_x + 2\eta_y + 2\zeta_z \nonumber \\
& + \left\langle d\xi, d\xi \right\rangle + \left\langle d\eta, d\eta \right\rangle + \left\langle d\zeta, d\zeta \right\rangle + 2\varphi(\eta_y-\xi_x) \label{eq:lambda1}
\end{align}
and
\begin{align}
\lambda_2  = & \,\, \tr\cof\left(\gamma^{ij}\right) \nonumber \\
= & \,\, 3 + 4\xi_x + 4\eta_y + 4\zeta_z + 4\xi_x\eta_y + 4\xi_x\zeta_z + 4\eta_y\zeta_z \nonumber \\
& + 2\left\langle d\xi, d\xi \right\rangle + 2\left\langle d\eta, d\eta \right\rangle + 2\left\langle d\zeta, d\zeta \right\rangle \nonumber \\
& - (\eta_x + \xi_y)^2 - (\zeta_x + \xi_z)^2 - ( \zeta_y + \eta_z)^2 + 2\varphi(\eta_y-\xi_x) \, \label{eq:lambda2}.
\end{align}
Note that, to this order,
\begin{equation} \label{linear}
\lambda_2 - 3 = (\lambda_0 - 1) + (\lambda_1 - 3)
\end{equation}
and
\begin{align} 
(\lambda_0 - 1)^2 & = (\lambda_1 - 3)^2 = \frac14 (\lambda_2 - 3)^2 = (\lambda_0 - 1)(\lambda_1 - 3) \nonumber \\
& = \frac12 (\lambda_0 - 1)(\lambda_2 - 3) = \frac12 (\lambda_1 - 3)(\lambda_2 - 3) \,  .\label{quadratic}
\end{align}
To second order, the elastic Lagrangian is then
\begin{align}
\mathcal{L} = & \,\,a_0 + a_1 (\lambda_0 - 1) + b_1 (\lambda_1 - 3) + c_1 (\lambda_2 - 3) \nonumber \\
& + a_2 (\lambda_0 - 1)^2 + b_2 (\lambda_1 - 3)^2 + c_2 (\lambda_2 - 3)^2 \nonumber \\
& + d_2 (\lambda_0 - 1) (\lambda_1 - 3) + e_2 (\lambda_0 - 1) (\lambda_2 - 3) + f_2 (\lambda_1 - 3) (\lambda_2 - 3) \, ,
\end{align}
or, using \eqref{linear} and \eqref{quadratic},
\begin{equation} \label{Lagrangian}
\mathcal{L} = a_0 + \tilde{a}_1 (\lambda_0 - 1) + \tilde{b}_1 (\lambda_1 - 3) + \tilde{a}_2 (\lambda_0 - 1)^2 \, ,
\end{equation}
where
\begin{align}
& \tilde{a}_1 = a_1 + c_1 \, , \\
& \tilde{b}_1 = b_1 + c_1 \, , \\
& \tilde{a}_2 = a_2 + b_2 + 4c_2 + d_2 + 2e_2 + 2f_2 \, .
\end{align}
As shown in \cite{Natario:2019nrf}, these coefficients can be related to the rest energy density $\rho_0$, the longitudinal speed of sound $c_L$ and the transverse speed of sound $c_T$ by the formulas
\begin{align}
\label{constants1} & \rho_0 = a_0 = 2\tilde{a}_1 + 2\tilde{b}_1 \, , \\
\label{constants2} & c_L^2 = \frac{\tilde{a}_1 + \tilde{b}_1 + 4\tilde{a}_2}{\tilde{a}_1 + \tilde{b}_1} \, , \\
\label{constants3} & c_T^2 = \frac{\tilde{b}_1}{\tilde{a}_1 + \tilde{b}_1} \, ,
\end{align}
so that the Lagrangian~\ref{Lagrangian} can be rewritten as
\begin{equation} \label{Lagrangian2}
\boxed{\mathcal{L} = \rho_0 \left[ 1 + \frac{1 - c_T^2}2 (\lambda_0 - 1) + \frac{c_T^2}2 (\lambda_1 - 3) + \frac{c_L^2 - 1}8 (\lambda_0 - 1)^2 \right] } \,\, .
\end{equation}
Note that the Lagrangian for an elastic solid in Minkowski spacetime can be obtained by setting $\varphi = 0$. Since $\lambda_0$ does not depend on $\varphi$, we see that \eqref{Lagrangian2} differs from the Lagrangian in Minkowski spacetime only by the interaction term
\begin{equation}
\mathcal{L}' = \rho_0c_T^2\varphi(\eta_y-\xi_x) \, .
\end{equation}
Since $\varphi=\varphi(t-z)$, we have
\begin{equation}
\partial_x \left( \frac{\partial\mathcal{L}' }{\partial \xi_x} \right) = \partial_y \left( \frac{\partial\mathcal{L}' }{\partial \eta_y} \right) = 0 \, ,
\end{equation}
and so the equations of motion are exactly the same as in Minkowski spacetime. In particular, $\xi\equiv\eta\equiv\zeta\equiv 0$ is a solution of the equations of motion. Unlike in Minkowski spacetime, however, this solution is not stress-free when a gravitational wave is present. In fact, from
\begin{equation} \label{Tmunu}
T^{\mu}_{\,\,\,\,\nu} = \frac{\partial\mathcal{L}}{\partial(\partial_\mu \bar{x}^i)} \partial_\nu \bar{x}^i - \mathcal{L} \delta^{\mu}_{\,\,\,\,\nu}
\end{equation}
(here we can use the canonical energy-momentum tensor as its covariant divergence coincides with its divergence to first order), we obtain, for the solution $\xi\equiv\eta\equiv\zeta\equiv 0$ (for which $d\bar{x}=dx$, $d\bar{y}=dy$ and $\mathcal{L'}=0$), the nonvanishing first order components
\begin{equation}
T^{x}_{\,\,\,\,x} = \frac{\partial\mathcal{L'}}{\partial\xi_x} \partial_x \bar{x} - \mathcal{L}' \delta^{x}_{\,\,\,\,x} = - \rho_0 c_T^2 \varphi
\end{equation}
and
\begin{equation}
T^{y}_{\,\,\,\,y} = \frac{\partial\mathcal{L'}}{\partial\eta_y} \partial_y \bar{y} - \mathcal{L}' \delta^{y}_{\,\,\,\,y} = \rho_0 c_T^2 \varphi \, .
\end{equation}
Notice that
\begin{equation}
\partial_x T^{x}_{\,\,\,\,x} = \partial_y T^{y}_{\,\,\,\,y} = 0 
\end{equation}
as $\varphi=\varphi(t-z)$. This again shows that the Euler-Lagrange equations are the same as in Minkowski, namely the Navier-Cauchy equations for linear elasticity of homogeneous and isotropic materials:
\begin{equation} \label{Navier-Cauchy}
\boxed{\partial^2_t \xi^{i} = c_T^2 \Delta \xi^i + (c_L^2 - c_T^2) \partial_i (\partial_j \xi^j)} \,\, .
\end{equation}
The gravitational wave will affect the equations of motion of the elastic body only through the boundary conditions for these equations.

As shown in \cite{Natario:2019nrf}  (after using \eqref{constants1}, \eqref{constants2} and \eqref{constants3}), the spatial components of the energy-momentum tensor when $\varphi \equiv 0$ are given by
\begin{equation} \label{TijMinkowski}
T^{ij} = \rho_0c_T^2 (\partial_i \xi^j + \partial_j \xi^i) + \rho_0(c_L^2 - 2c_T^2) (\partial_k \xi^k) \delta_{ij} \, ,
\end{equation}
and so in the general case we have
\begin{equation} \label{Tij}
T^{ij} = \rho_0c_T^2 (\partial_i \xi^j + \partial_j \xi^i) + \rho_0(c_L^2 - 2c_T^2) (\partial_k \xi^k) \delta_{ij} - \rho_0 c_T^2 \varphi \delta_{i1}\delta_{j1} + \rho_0 c_T^2 \varphi \delta_{i2}\delta_{j2} \, .
\end{equation}
If $n^i$ is the unit normal to the boundary of the elastic body then the natural boundary condition $T^{ij} n^j = 0$ reads
\begin{equation} \label{boundary}
\boxed{c_T^2 (\partial_i \xi^j + \partial_j \xi^i) n^j + (c_L^2 - 2c_T^2) (\partial_k \xi^k) n^i - c_T^2 \varphi \delta_{i1} n^1 + c_T^2 \varphi \delta_{i2} n^2 = 0} \, .
\end{equation}
Therefore, the gravitational wave modifies the boundary conditions whenever the body does not behave like a perfect fluid, that is, whenever $c_T \neq 0$. In particular, $\xi^i=0$ will not satisfy the boundary conditions in the presence of a gravitational wave, and so a solid initially in equilibrium will be set into motion by the gravitational wave and continue to oscillate after the wave has passed.\footnote{By contrast, $\xi^i=0$ is a solution for a material behaving as a perfect fluid ($c_T=0$). Consequently, although the gravitational wave modifies the distances between neighboring particles while it is present, an initially relaxed body experiences no permanent excitation and returns to rest once the wave has passed.
}

Finally, it is interesting to note that if we write 
\begin{equation}
h = g - \eta = \varphi(t-z) dx^2 - \varphi(t-z) dy^2 \, , 
\end{equation}
where $\eta$ denotes the Minkowski metric, and if $T^{\mu\nu}$ is the energy-momentum tensor \eqref{TijMinkowski} for the elastic material in Minkowski spacetime, then the interaction term can be written as
\begin{equation}
\mathcal{L}'= -\frac12 T^{xx} \varphi + \frac12 T^{yy} \varphi = -\frac12 T^{\mu\nu} h_{\mu\nu} \, ,
\end{equation}
which is precisely Dyson's interaction Lagrangian in the effective potential approach \cite{Dyson:1969zgf}.
%
%
\section{Elastic response of a rectangular plate to a gravitational wave}\label{section3}
We now apply the linearized equations of motion to the problem of a thin rectangular plate under the influence of a passing gravitational wave. In order to obtain explicit expressions, we assume that the plate is orthogonal to the direction of propagation of the wave and aligned with its polarization.

\begin{figure}[h]
    \centering
    \includegraphics[width=0.75\linewidth]{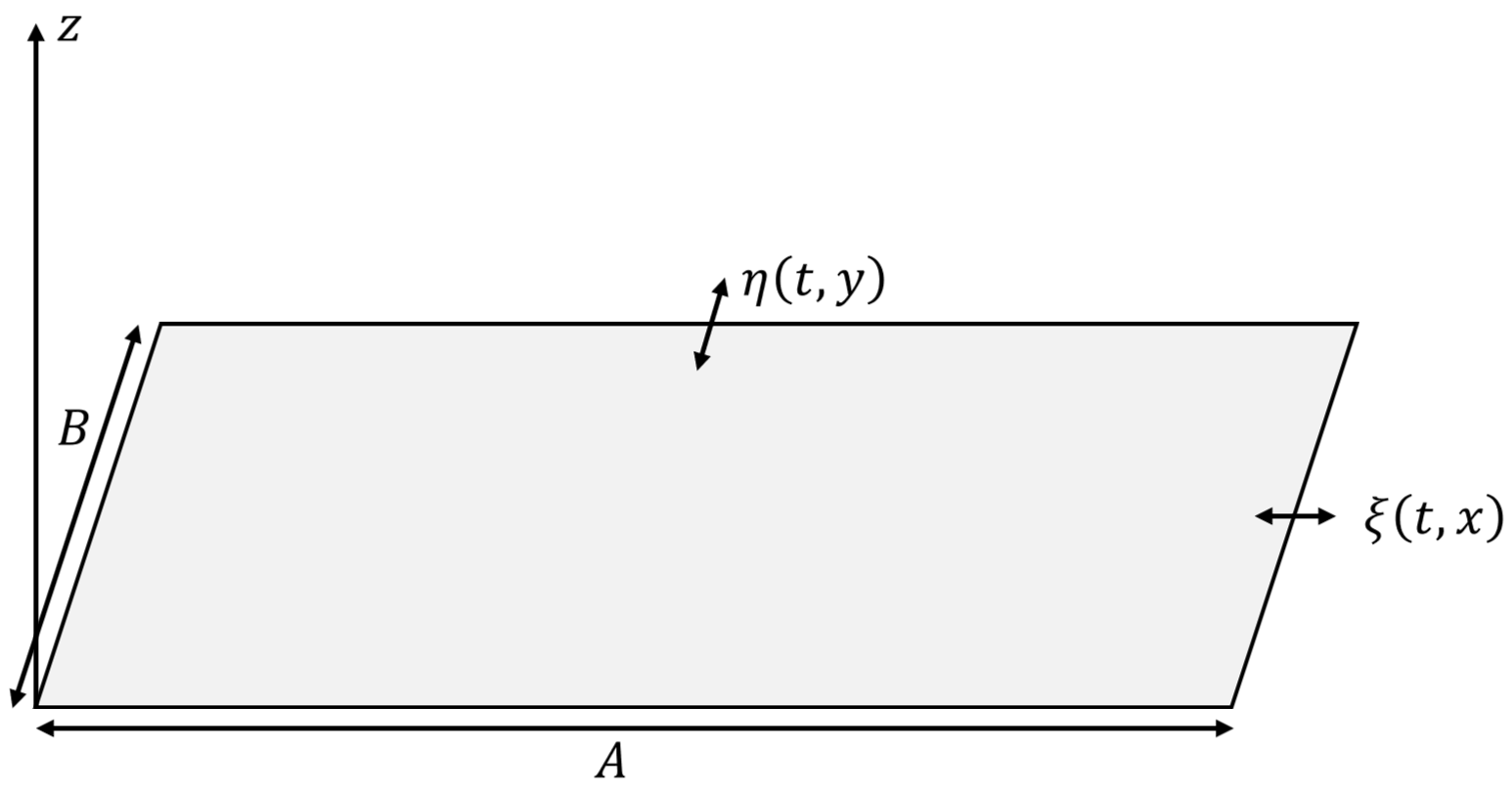}
    \caption{Rectangular plate of dimensions $A$ and $B$ lying in the plane $z=0$. The functions $\xi(t,x)$ and $\eta(t,y)$ represent minus the displacements in the $x$ and $y$ directions, respectively.}
    \label{fig:placeholder}
\end{figure}

Let us then consider a rectangular plate, occupying the region of space defined by the conditions
\begin{equation}
0 \leq x \leq A \, , \qquad 0 \leq y \leq B \, , \qquad -\delta  \leq z \leq 0 \, ,
\end{equation}
where $0 < \delta \ll A,B$. We will adopt the ansatz
\begin{equation} \label{ansatz}
\xi = \xi(t,x) \, , \qquad \eta = \eta(t,y) \, , \qquad \zeta \equiv 0 \, ,
\end{equation}
under which the Navier–Cauchy equations decouple to yield
\begin{equation} \label{wave_equations}
\boxed{\ddot{\xi} = c_L^2 \xi''} \, , \qquad \qquad \boxed{\ddot{\eta} = c_L^2 \eta''} \, ,
\end{equation}
where $\dot{} \equiv \partial_t$ and $'\equiv \partial_x$ or $'\equiv \partial_y$ as appropriate.
The boundary conditions on the face $z=0$ reduce to the single equation
\begin{equation}
(c_L^2 - 2c_T^2) (\xi'(t,x) + \eta'(t,y)) = 0 \, ,
\end{equation}
which is of course impossible for nontrivial $\xi(t,x)$ and $\eta(t,y)$ unless
\begin{equation}
\boxed{c_L^2 = 2c_T^2 \Leftrightarrow \nu = 0} \, ,
\end{equation}
where $\nu$ is the Poisson ratio (see \cite{Natario:2019nrf}). We henceforth assume that this condition holds, which also automatically enforces the boundary conditions on the face $z=-\delta$. The boundary conditions on the face $x=A$ then reduce to the single equation
\begin{equation}
2c_T^2 \xi'(t,A) - c_T^2 \varphi(t-z) = 0 \, ,
\end{equation}
Since we must assume $c_T^2 \neq 0 \Leftrightarrow c_L^2 \neq 0$, we see that this condition is again impossible unless we take the limit $\delta \to 0$, in which case it becomes simply
\begin{equation} \label{boundaryxA}
\boxed{\xi'(t,A) = \frac12 \varphi(t)} \, .
\end{equation}
In this limit, the boundary conditions on the face $x=0$ become
\begin{equation} \label{boundaryx0}
\boxed{\xi'(t,0) = \frac12 \varphi(t)} \, ,
\end{equation}
and similarly
\begin{equation} \label{boundaryyB}
\boxed{\eta'(t,B) = - \frac12 \varphi(t)}
\end{equation}
on the face $y=B$ and
\begin{equation} \label{boundaryy0}
\boxed{\eta'(t,0) = - \frac12 \varphi(t)}
\end{equation}
on the face $y=0$. Thus we have reduced the Navier-Cauchy equations to the two decoupled one-dimensional wave equations \eqref{wave_equations} with the Neumann boundary conditions~\eqref{boundaryxA}--\eqref{boundaryy0}.

Let us now compute the total energy of the plate. Using the ansatz~\eqref{ansatz}, we can simplify equations \eqref{eq:lambda0} and \eqref{eq:lambda1} to obtain
\begin{equation}
\lambda_0  = 1 + 2\xi' + 2\eta' + 4\xi'\eta' - \dot{\xi}^2 + \xi'^2 - \dot{\eta}^2 + \eta'^2
\end{equation}
and
\begin{align}
\lambda_1 = 3 + 2\xi' + 2\eta' - \dot{\xi}^2 + \xi'^2 - \dot{\eta}^2 + \eta'^2 + 2\varphi(\eta'-\xi') \, ,
\end{align}
and so \eqref{Lagrangian2}, with $c_L^2 = 2c_T^2$, yields
\begin{equation}
\mathcal{L} = \rho_0 \left[ 1 + \xi' + \eta' + \xi'\eta' - \frac12\dot{\xi}^2 - \frac12\dot{\eta}^2 + \frac12 c_L^2 \left( \xi'^2 + \eta'^2 + \varphi(\eta'-\xi')\right) \right] \, . \label{eq:elastic_law}
\end{equation}
The canonical energy density is then
\begin{align}
T_{00} & = - \frac{\partial\mathcal{L}}{\partial \dot{\xi}} \dot{\xi} - \frac{\partial\mathcal{L}}{\partial \dot{\eta}} \dot{\eta} + \mathcal{L} \nonumber \\
& = \rho_0 \left[ (1 + \xi')(1 + \eta') + \frac12\dot{\xi}^2 + \frac12\dot{\eta}^2 + \frac12 c_L^2 \left( \xi'^2 + \eta'^2 + \varphi(\eta'-\xi')\right) \right] \, .
\end{align}
Since $\xi^i$ can be seen as minus the displacement, and the volume element of the spatial part of the metric \eqref{metricGW} is simply $1$ to linear order, the total energy of the plate as measured by observers with constant spatial coordinates is
\begin{equation}
E(t) = \int_{-\xi(t,0)}^{A-\xi(t,A)} \int_{-\eta(t,0)}^{B-\eta(t,B)} \int_{-\delta}^0 T_{00} \, dz \, dy \, dx \, .
\end{equation}
Noting that, to quadratic order,
\begin{align}
\int_{-\xi(t,0)}^{A-\xi(t,A)} (1 + \xi') \, dx & = A-\xi(t,A)+\xi(t,0)+\xi(t,A-\xi(t,A))-\xi(t,-\xi(t,0)) \nonumber \\
& = A-\xi(t,A)+\xi(t,0)+\xi(t,A)-\xi'(t,A)\,\xi(t,A)-\xi(t,0)+\xi'(t,0)\,\xi(t,0) \nonumber \\
& = A - \frac12 \varphi(t) \, \xi(t,A) + \frac12 \varphi(t) \, \xi(t,0) \, ,
\end{align}
and similarly
\begin{equation}
\int_{-\eta(t,0)}^{B-\eta(t,B)} (1 + \eta') \, dy = B + \frac12 \, \varphi(t) \eta(t,B) - \frac12 \varphi(t) \, \eta(t,0) \, ,
\end{equation}
we have
\begin{align}
E(t) = \sigma_0 AB & + \frac12 \sigma_0 (1+c_L^2) \varphi(t)\Big[B\big(\xi(t,0)-\xi(t,A)\big)+A\big(\eta(t,B)-\eta(t,0)\big)\Big] \nonumber \\
& \qquad \qquad \qquad + \frac12 \sigma_0 \int_{0}^{B} \int_{0}^{A}  \left[ \dot{\xi}^2 + \dot{\eta}^2 + c_L^2 \left( \xi'^2 + \eta'^2\right) \right] dx \, dy \, ,
\end{align}
where $\sigma_0 = \rho_0 \delta$ is the rest energy density per unit area. In particular, before or after the gravitational wave has passed (that is, when $\varphi(t)= 0$), we have\footnote{It is easily verified that \(E(t)\) is conserved for solutions of the wave equations~\eqref{wave_equations} satisfying the boundary conditions~\eqref{boundaryxA}--\eqref{boundaryy0} whenever \(\varphi(t)=0\). In particular, if the plate is initially at rest and subsequently interacts with a gravitational wave, then the absorbed energy is given by \(\Delta E = E(t)-\sigma_0 AB\), where \(t\) denotes any instant after the wave has passed.}
\begin{equation} \label{energy}
\boxed{E(t) = \sigma_0 AB + \frac12 \sigma_0 \int_{0}^{B} \int_{0}^{A}  \left[ \dot{\xi}^2 + \dot{\eta}^2 + c_L^2 \left( \xi'^2 + \eta'^2\right) \right] dx \, dy} \, .
\end{equation}

Finally, let us determine an explicit expression for the solution to the plate's equations of motion. To solve the initial boundary value problem for $\xi$, given by the wave equation in~\eqref{wave_equations} together with the Neumann boundary conditions \eqref{boundaryxA}--\eqref{boundaryx0}, we start by noting that it implies that $\xi'$ is also a solution of the wave equation with speed $c_L$, and so
\begin{equation}
\xi'(t,x) = f(c_L t - x) + g(c_L t + x) 
\end{equation}
for appropriate smooth functions $f$ and $g$. Imposing the first boundary condition yields
\begin{equation}
\xi'(t,0) = \frac12 \varphi(t) \Leftrightarrow f(c_L t) + g(c_L t) = \frac12 \varphi(t) \Leftrightarrow g(u) = -f(u) + \frac12 \varphi\left(\frac{u}{c_L}\right) \, ,
\end{equation}
whence
\begin{equation}
\xi'(t,x) = f(c_L t - x) - f(c_L t + x) + \frac12 \varphi\left(\frac{c_L t + x}{c_L}\right) \, ,
\end{equation}
and so imposing the second boundary condition gives
\begin{align}
\xi'(t,A) = \frac12 \varphi(t) & \Leftrightarrow f(c_L t - A) - f(c_L t + A) + \frac12 \varphi\left(\frac{c_L t + A}{c_L}\right) = \frac12 \varphi(t) \nonumber \\
& \Leftrightarrow f(u) = f(u-2A) + \frac12 \varphi\left(\frac{u}{c_L}\right) - \frac12 \varphi\left(\frac{u - A}{c_L}\right) \, .
\end{align}
Iterating this identity, and assuming that
\begin{equation}
\lim_{u\to-\infty}f(u) = 0 \, , 
\end{equation}
that is, that the plate is at rest in the remote past, leads to
\begin{align}
f(u) & = f(u-4A) + \frac12 \varphi\left(\frac{u - 2A}{c_L}\right) - \frac12 \varphi\left(\frac{u - 3A}{c_L}\right) + \frac12 \varphi\left(\frac{u}{c_L}\right) - \frac12 \varphi\left(\frac{u - A}{c_L}\right) = \dots \nonumber \\
& = \frac12 \sum_{n=0}^{+\infty} (-1)^n\varphi\left(\frac{u - nA}{c_L}\right) \, ,
\end{align}
and consequently to
\begin{equation} \label{xiprime}
\boxed{\xi'(t,x) = \frac12 \sum_{n=0}^{+\infty} (-1)^n\varphi\left(\frac{c_L t - x - nA}{c_L}\right) + \frac12 \sum_{n=1}^{+\infty} (-1)^{n+1}\varphi\left(\frac{c_L t + x - nA}{c_L}\right)} \, .
\end{equation}
Setting
\begin{equation} \label{Phi(u)}
\Phi(u) = \int_{-\infty}^u \varphi(s) ds \, , 
\end{equation}
we have
\begin{equation} \label{xi}
\boxed{\xi(t,x) = \frac{c_L}2 \sum_{n=0}^{+\infty} (-1)^{n+1} \Phi\left(\frac{c_L t - x - nA}{c_L}\right) + \frac{c_L}2 \sum_{n=1}^{+\infty} (-1)^{n+1} \Phi\left(\frac{c_L t + x - nA}{c_L}\right)} \, ,
\end{equation}
whence
\begin{equation} \label{xidot}
\boxed{\dot{\xi}(t,x) = \frac{c_L}2 \sum_{n=0}^{+\infty} (-1)^{n+1} \varphi\left(\frac{c_L t - x - nA}{c_L}\right) + \frac{c_L}2 \sum_{n=1}^{+\infty} (-1)^{n+1} \varphi\left(\frac{c_L t + x - nA}{c_L}\right)} \, .
\end{equation}
It can be shown that these results agree with those obtained in \cite{Vicente:2024ksp} for a one-dimensional rod, using very different methods.\footnote{A one-dimensional rod can be seen as the limit case $B \to 0$, but it was studied in \cite{Vicente:2024ksp} directly from the equations of motion for a one-dimensional elastic object.} The corresponding formulae for $\eta'(t,x)$, $\eta(t,x)$ and $\dot{\eta}(t,x)$ are readily obtained by making the substitutions $x\mapsto y$, $A\mapsto B$ and $\varphi\mapsto-\varphi$.
%
%
\section{Example: short gravitational wave burst}\label{section4}
Let us now compute the plate's reaction to different types of gravitational wave signals. As a first example, consider a gravitational wave burst whose duration is shorter than half the minimum sound-crossing time of the plate. Accordingly, we assume that the support of $\varphi$ satisfies
\begin{equation} \label{supp}
\supp \varphi \subset \left[ -\frac{L}{2c_L},0 \right] \, ,
\end{equation}
where $L=\min\{A,B\}$. Then, for $t=0$, each series in the expansions \eqref{xiprime} and \eqref{xidot} reduces to a single term, yielding, for $x \in [0,A]$,
\begin{equation}
\xi'\left(0,x\right) = \frac12 \varphi\left(-\frac{x}{c_L}\right) + \frac12 \varphi\left(\frac{x-A}{c_L}\right)
\end{equation}
and
\begin{equation}
\dot{\xi}\left(0,x\right) = - \frac{c_L}2 \varphi\left(-\frac{x}{c_L}\right) + \frac{c_L}2 \varphi\left(\frac{x-A}{c_L}\right) \, ,
\end{equation}
with similar expressions for $\eta'\left(0,y\right)$ and $\dot{\eta}\left(0,y\right)$. Since the supports of the two functions in each of these expressions do not overlap, we obtain from \eqref{energy}
\begin{align}
E(0) & = \sigma_0 AB + \frac12 \sigma_0 B \int_{0}^{A} \left( 2 \times \frac{c_L^2}4 \varphi^2\left(-\frac{x}{c_L}\right) + 2 \times \frac{c_L^2}4 \varphi^2\left(\frac{x-A}{c_L}\right) \right) \nonumber dx \\ 
& \qquad \qquad +\frac12 \sigma_0 A \int_{0}^{B} \left( 2 \times \frac{c_L^2}4 \varphi^2\left(-\frac{y}{c_L}\right) + 2 \times \frac{c_L^2}4 \varphi^2\left(\frac{y-B}{c_L}\right) \right) dy \\
 & = \sigma_0 AB + \sigma_0 c_L^3 \frac{A+B}{2} \int_\bbR \varphi^2(t) dt \, . \nonumber
\end{align}
In other words, the energy absorbed by the plate is
\begin{equation} \label{DeltaEburst}
\boxed{
\Delta E = \sigma_0 c_L^3 \frac{A+B}{2} \int_\bbR \varphi^2(t) dt} \, .
\end{equation}
Let us compare this with the total gravitational wave energy incident on the plate. The energy density of the gravitational wave is \cite[pp.~955]{Misner:1973prb}
\begin{equation}
T_{00}^{\rm (GW)} = \frac1{32\pi} \left\langle \partial_t h_{ij}^{\rm TT} \partial_t h_{ij}^{\rm TT} \right\rangle = \frac1{16\pi} \left(\varphi'(t-z)\right)^2 \, ,
\label{eq:GWPower}
\end{equation}
and so the total energy incident on the rectangular plate is
\begin{equation}
E^{\rm (GW)} = \frac{AB}{16\pi} \int_{\bbR} \left(\varphi'(z)\right)^2 dz \, .
\end{equation}
We can compare $\Delta E$ with $E^{\rm (GW)}$ by using Poincar\'e's inequality for compactly supported smooth functions, which in this case can be deduced from \eqref{supp}: since
\begin{align}
\int_{\bbR} \varphi^2(z) dz & = \int_{\bbR} z' \varphi^2(z) dz = - 2 \int_{\bbR} z \varphi(z) \varphi'(z) dz \leq 2 \int_{\bbR} \frac{L}{2c_L} \left|\varphi(z)\right| \left|\varphi'(z)\right| dz \nonumber \\
& \leq \frac{L}{c_L} \left( \int_\bbR \varphi^2(z) dz \right)^\frac12 \left( \int_\bbR \left(\varphi'(z)\right)^2 dz \right)^\frac12 \, ,
\end{align}
where we used the Cauchy-Schwarz inequality, we have
\begin{equation}
\int_{\bbR} \varphi^2(z) dz \leq \frac{L^2}{c_L^2} \int_\bbR \left(\varphi'(z)\right)^2 dz \, .
\end{equation}
Therefore,
\begin{align}
\Delta E & \leq \sigma_0 L^2 c_L \frac{A+B}{2} \int_\bbR \left(\varphi'(z)\right)^2 dz \nonumber \\
    & = 16 \pi \sigma_0 L^2 c_L \frac{A+B}{2AB} E^{\rm (GW)} \leq \frac{16 \pi Mc_L}{L}E^{\rm (GW)} \, ,
\end{align}
where $M=\sigma_0AB$ is the total rest mass of the plate, and we used the fact that $\frac{2AB}{A+B} \geq L$. In our test body approximation, we must of course have
\begin{equation}
\frac{M}{L} \ll 1\, ,
\end{equation}
and, since $c_L \leq 1$, we see that
\begin{equation}
\boxed{\Delta E \ll E^{\rm (GW)}} \, ,
\end{equation}
that is, the energy deposited on the plate is much smaller than the total incident energy of the gravitational wave.
%
%
\section{Example: continuous harmonic gravitational wave}\label{section5}

As a second example, let us take
\begin{equation}
\varphi(t) = 
\begin{cases}
\amplitude \sin\left(\omega t \right) \text{ if } t \leq 0 \\
0 \text{ otherwise }
\end{cases}
\, ,
\end{equation}
corresponding to an infinitely long harmonic gravitational wave signal with amplitude $\amplitude$ and angular frequency $\omega$, which terminates at time $t=0$. Since the duration of a physically realistic signal is expected to be much longer than the vibration periods of the plate, but much shorter than the relaxation time due to internal dissipation (which is not included in our model), we will use Abel regularization (which ``forgets'' about very early times) to sum the series and integrals that might diverge due to the infinite duration of the signal. We have
\begin{align}
\xi'(0,x) & = \frac12 \sum_{n=0}^{+\infty} (-1)^n\amplitude\sin\left(\frac{\omega(- x - nA)}{c_L}\right) + \frac12 \sum_{n=1}^{+\infty} (-1)^{n+1}\amplitude\sin\left(\frac{\omega(x - nA)}{c_L}\right) \nonumber \\
& = - \frac12 \sum_{n=0}^{+\infty} \amplitude\sin\left(\frac{\omega x}{c_L} + n \left( \pi + \frac{\omega A}{c_L} \right) \right) - \frac12 \sum_{n=1}^{+\infty} \amplitude\sin\left(\frac{\omega x}{c_L} + n \left( \pi - \frac{\omega A}{c_L} \right) \right)
\end{align}
and
\begin{align}
\dot{\xi}(0,x) & = \frac{c_L}2 \sum_{n=0}^{+\infty} (-1)^{n+1} \amplitude\sin\left(\frac{\omega(- x - nA)}{c_L}\right) + \frac{c_L}2 \sum_{n=1}^{+\infty} (-1)^{n+1} \amplitude\sin\left(\frac{\omega(x - nA)}{c_L}\right) \nonumber \\
& = \frac{c_L}2 \sum_{n=0}^{+\infty} \amplitude\sin\left(\frac{\omega x}{c_L} + n \left( \pi + \frac{\omega A}{c_L} \right) \right) - \frac{c_L}2 \sum_{n=1}^{+\infty} \amplitude\sin\left(\frac{\omega x}{c_L} + n \left( \pi - \frac{\omega A}{c_L} \right) \right) \, .
\end{align}
From the partial sum expression
\begin{equation}
\sum_{n=1}^N \sin \left(n \alpha + \beta \right) = \Imag \sum_{n=1}^N e^{i (n \alpha + \beta)} = \Imag \left( e^{i (\alpha + \beta)} \frac{1 - e^{i N \alpha}}{1 - e^{i \alpha}} \right)
\end{equation}
it is clear that the series
\begin{equation}
\sum_{n=1}^{+ \infty} \sin \left(n \alpha + \beta \right) 
\end{equation}
diverges.\footnote{This reflects the fact that the wave interacts with the plate over the infinite interval $t\in(-\infty,0)$ in the absence of dissipation.} However, we can still obtain a solution by noting that the series is Abel-summable since, for $\alpha \not\in 2\pi \bbZ$,
\begin{align}
\lim_{\varepsilon \to 0^+} \sum_{n=1}^{+ \infty} \sin \left(n \alpha + \beta \right) e^{-\varepsilon n} & = \lim_{\varepsilon \to 0^+} \Imag \sum_{n=1}^{+ \infty} e^{(i \alpha -\varepsilon) n + i \beta} = \lim_{\varepsilon \to 0^+} \Imag \frac{e^{i (\alpha + \beta) - \varepsilon}}{1 - e^{i \alpha - \varepsilon}} \nonumber \\
& = \Imag \frac{e^{i (\alpha + \beta)}}{1 - e^{i \alpha}} = \frac{\sin(\alpha + \beta) - \sin \beta}{2(1 - \cos \alpha)} \, .
\end{align}
If we then take
\begin{equation} \label{Abelsumsin}
\sum_{n=1}^{+ \infty} \sin \left(n \alpha + \beta \right) = \frac{\sin(\alpha + \beta) - \sin \beta}{2(1 - \cos \alpha)} \Rightarrow \sum_{n=0}^{+ \infty} \sin \left(n \alpha + \beta \right) = \frac{\sin(\alpha - \beta) + \sin \beta}{2(1 - \cos \alpha)}
\end{equation}
we have, for $\frac{\omega A}{c_L} \not\in \pi + 2\pi\bbZ$,
\begin{equation} \label{xiprime(0)}
\xi'(0,x) = - \frac{\amplitude}4 \, \frac{\sin\left(\pi + \frac{\omega A}{c_L} - \frac{\omega x}{c_L} \right) + \sin\left(\frac{\omega x}{c_L}\right) }{1 - \cos\left(\pi + \frac{\omega A}{c_L}\right) } - \frac{\amplitude}4 \, \frac{\sin\left(\pi - \frac{\omega A}{c_L} + \frac{\omega x}{c_L} \right) - \sin\left(\frac{\omega x}{c_L}\right) }{1 - \cos\left(\pi - \frac{\omega A}{c_L}\right) } = 0
\end{equation}
and analogously
\begin{equation} \label{dotxi(0)}
\dot{\xi}(0,x) = \frac{\amplitude c_L}2 \, \frac{\sin\left(\frac{\omega x}{c_L} - \frac{\omega A}{c_L} \right) + \sin\left(\frac{\omega x}{c_L}\right) }{1 + \cos\left(\frac{\omega A}{c_L}\right) } = \frac{\amplitude c_L \sin\left(\frac{\omega x}{c_L} - \frac{\omega A}{2c_L} \right)}{2\cos\left(\frac{\omega A}{2c_L}\right) } \, ,
\end{equation}
with similar expressions for $\eta'(0,y)$ and $\dot{\eta}(0,y)$, where we should make the substitutions $A\mapsto B$, $x\mapsto y$ and change the overall sign. Therefore, we have from \eqref{energy}
\begin{align}
E(0) & = \sigma_0 AB + \frac12 \sigma_0 \int_{0}^{B}\int_{0}^{A} \left( \dot \xi^2(0,x) + \dot \eta^2(0,y) \right) dx \, dy \nonumber \\
& = \sigma_0 AB + \sigma_0\frac{ABc_L^2\amplitude^2}{16}\left[ \frac{1}{\cos^2(\frac{\omega A}{2c_L})}+\frac{1}{\cos^2(\frac{\omega B}{2c_L})} - \frac{2c_L}{\omega A}\tan\left(\frac{\omega A}{2c_L}\right)- \frac{2c_L}{\omega B}\tan\left(\frac{\omega B}{2c_L}\right) \right] \, ,
\end{align}
that is, the total energy absorbed by the plate is
\begin{equation} \label{DeltaEcontinuous}
\boxed{\Delta E = \frac{Mc_L^2\amplitude^2}{16}\Big[2+\tan^2 \left(\theta_A\right)+\tan^2 \left(\theta_B\right)- \tan\left(\theta_A\right)/\theta_A - \tan\left(\theta_B\right)/\theta_B \Big]} \, ,
\end{equation}
where
\begin{equation}
\theta_{A} = \frac{\omega A}{2c_L} \qquad \text{ and } \qquad \theta_{B} = \frac{\omega B}{2c_L} \, .
\end{equation}
Note that this expression becomes singular for
\begin{equation}
\theta_A \in \frac{\pi}2 + \pi\bbZ \qquad \text{ or } \qquad \theta_B \in \frac{\pi}2 + \pi\bbZ  \, ,
\end{equation}
that is, when the frequency of the gravitational wave coincides with one of the longitudinal normal mode frequencies of the plate. This resonant divergence is to be expected, since our model does not include any damping mechanism. It is also interesting to note that, as the frequency of the incident wave goes to $0$ (adimensionally, as $\theta_A, \theta_B \to 0$), the absorbed energy takes the form
\begin{equation}
    \Delta E \approx \frac{M c_L^2 h^2}{16}\cdot \frac{2}{3}\left(\frac{\omega^2 A^2}{4c_L^2}+\frac{\omega^2 B^2}{4c_L^2}\right) = \frac{M h^2 \omega^2}{96}\left(A^2+B^2\right).
\end{equation}
The expression above does not depend on $c_L$, showing that for low enough frequencies the gravitational wave no longer produces elastic waves on the plate, and it goes to $0$ as $\omega\to 0$, as expected.
%
%
%
\section{Gravitational wave emission}\label{section6}
\begin{figure}[h]
    \centering
    \includegraphics[width=.8\linewidth]{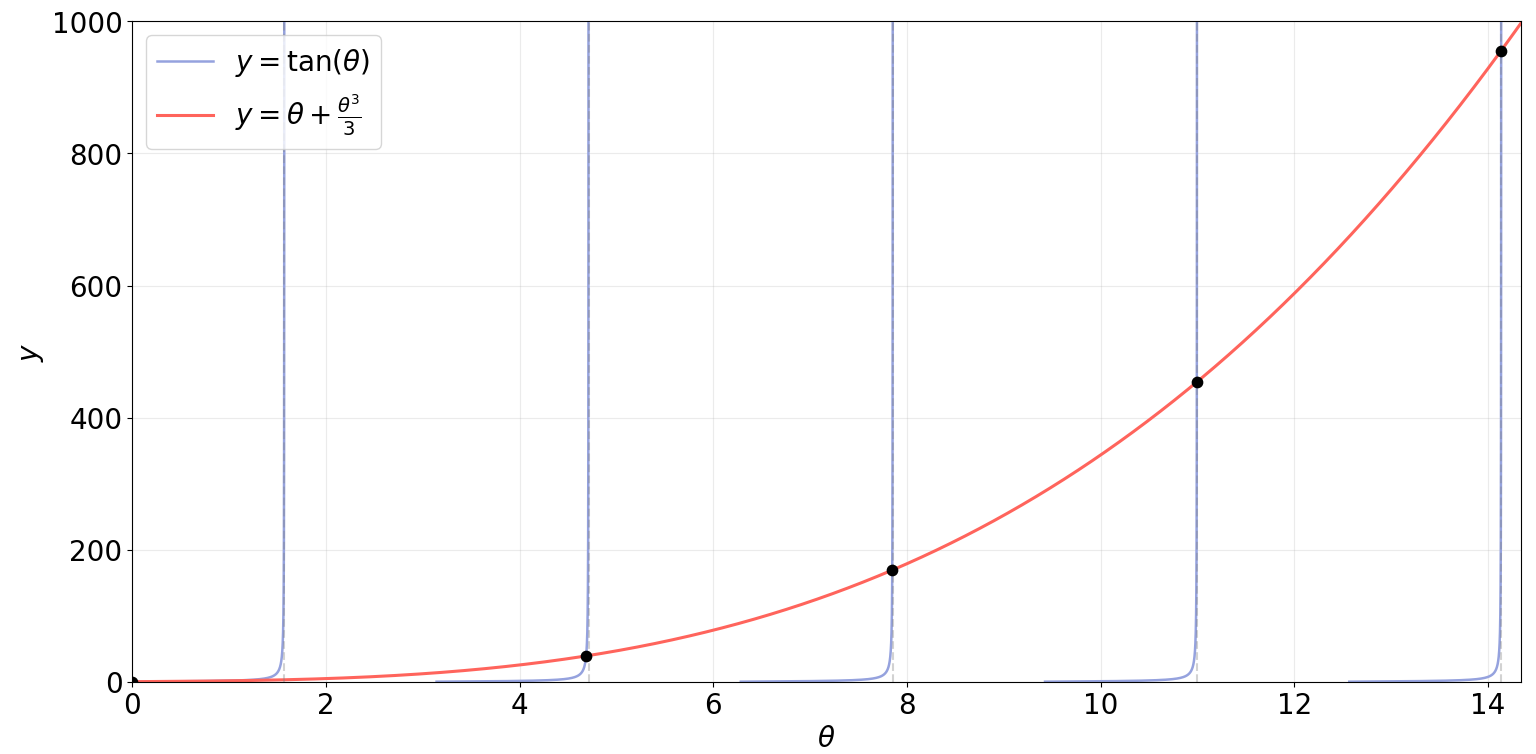}
    \caption{Plots for $y=\tan(\theta)$ and $y = \theta + \theta^3/3$. The intersection points of these plots are the roots of $g(\theta)$. As the figure clearly shows, they rapidly approach $\frac{2n+1}{2}\pi$ as $\theta$ increases.}
    \label{fig:intersections}
\end{figure}

We conclude by analyzing the gravitational wave emission of the rectangular plate when driven by the continuous harmonic signal introduced in Section~\ref{section5}. We first compute the displacement. Since the integral in \eqref{Phi(u)} diverges, we define it by Abel regularization:
\begin{equation} 
\Phi(u) = \lim_{\varepsilon \to 0} \int_{-\infty}^u \amplitude \sin(\omega s) e^{\varepsilon s} ds = \Imag \lim_{\varepsilon \to 0} \int_{-\infty}^u \amplitude e^{\varepsilon s + i \omega s} ds = \amplitude \Imag \frac{e^{i \omega u}}{i \omega} = - \frac{\amplitude}{\omega} \cos(\omega u)\, . 
\end{equation}
It is then straightforward to use \eqref{xi}, together with the cosine version of \eqref{Abelsumsin},
\begin{equation} 
\sum_{n=1}^{+ \infty} \cos \left(n \alpha + \beta \right) \equiv \frac{\cos(\alpha + \beta) - \cos \beta}{2(1 - \cos \alpha)} \Rightarrow \sum_{n=0}^{+ \infty} \cos \left(n \alpha + \beta \right) = \frac{\cos \beta - \cos(\alpha - \beta)}{2(1 - \cos \alpha)} \, ,
\end{equation}
to show that\footnote{Note the consistency with \eqref{xiprime(0)}, \eqref{dotxi(0)}.}
\begin{equation}
    \boxed{\xi(t,x)=\frac{\amplitude c_L}{2\omega}\  \frac{\sin\left(\frac{\omega (x-A/2)}{c_L}\right)}{\cos(\theta_A)}\ \sin{(\omega t)}} \, , \label{eq:displacement_x_sinusoidal}
\end{equation}
as well as the analogous formula
\begin{equation}
    \boxed{\eta(t,y)=-\frac{\amplitude c_L}{2\omega} \ \frac{\sin\left(\frac{\omega (y-B/2)}{c_L}\right)}{\cos(\theta_B)}\ \sin{(\omega t)}} \, .
\end{equation}
Note that the center of mass of the plate remains stationary. In order to compute the quadrupole moment, we take the elastic law (Eq.~\ref{eq:elastic_law}) and truncate it to linear order, yielding
\begin{equation}
    \rho = \rho_0\left(1+\xi'\right)\left(1+\eta'\right) \, .
\end{equation}
The mass element is then given by
\begin{equation}
    \rho \, dx \, dy \, dz = \rho_ 0\left(1+\xi'\right)\left(1+\eta'\right) dx \, dy \, dz = \rho_0 \, d\bar x \, d\bar y \, dz,
    \label{eq:massElement}
\end{equation}
where $\bar x \in [0,A]$ and $\bar y \in [0,B]$. Since the transverse-traceless ($TT$) frame is not locally inertial, we cannot use it to compute the quadrupole moment. Instead, we define the local inertial coordinates around the center of mass, given (to a good enough approximation) by
\begin{equation}
    \begin{cases}
        X = (1+\varphi/2) (x-A/2) = (1+\varphi/2) (\bar x - \xi -A/2), \\
        Y = (1-\varphi/2) (y-B/2) = (1-\varphi/2) (y - \eta -B/2).
    \end{cases}
    \label{eq:localInertFram}
\end{equation}
We now apply the quadrupole formula \cite{Maggiore:2007ulw}
\begin{equation}
M_{ij}=\int_{\rm plate} \rho \, X^i X^j d^3X \, ,
\end{equation}
together with Eqs.~\eqref{eq:massElement} and \eqref{eq:localInertFram}, to obtain
\begin{align}
    M_{xx}&=\sigma_0 B(1+\varphi)\int_{0}^{A}(\bar x-\xi-A/2)^2 d\bar x = M\left[\frac{A^2(1+\varphi)}{12}-\frac{2}{A}\int_{0}^{A}(\bar x-A/2) \xi \, d\bar x\right] \nonumber \\
    & = M\left[\frac{A^2(1+\varphi)}{12}-\varphi\frac{c_L^2}{\omega^2}\left(\frac{\tan(\theta_A)}{\theta_A}-1\right)\right] \nonumber \\ &= M\left[\frac{A^2}{12}+\varphi\left(\frac{A^2}{12}+\frac{c_L^2}{\omega^2}\left(1-\frac{\tan(\theta_A)}{\theta_A}\right)\right)\right].
\end{align}
This in turn yields
\begin{equation} \label{Mxx}
    \ddot M_{xx} = Mc_L^2 ~g(\theta_A)~h\sin(\omega t) \, ,
\end{equation}
where
\begin{equation}
    g(\theta) := \frac{\tan(\theta)}{\theta} - 1-\frac{\theta^2}{3} \, .
\end{equation}
Note that $g(\theta)$ is obtained from $\tan(\theta)/\theta$ by subtracting the first two terms in its Taylor expansion, so $g(\theta) \approx \frac{2}{15}\theta^4$ as $\theta \to 0$. An analogous expression for $M_{yy}$ is obtained by substituting $\xi \to \eta$, $A \to B$ and $\varphi \to -\varphi$ and, by symmetry, it is clear that $M_{xy}=M_{zz}=0$. Finally, we can compute the re-emitted wave along the $z$ axis, which is given by
\begin{equation}
    h^{TT}_{xx} = -h^{TT}_{yy} = \frac{2}{r}\frac{\ddot M_{xx} - \ddot M_{yy}}{2}=\frac{M c_L^2}{r}\Big(g(\theta_A)+g(\theta_B)\Big) h \sin(\omega t) \, .
    \label{eq:GWemissionZ}
\end{equation}
Remarkably, the rectangular plate emits no gravitational radiation back in the $z$ direction when $g(\theta_A)=-g(\theta_B)$, that is, when
\begin{equation}
    \frac{c_L}{A\omega} \tan\left(\frac{A\omega}{2c_L}\right)+\frac{c_L}{B\omega} \tan\left(\frac{B\omega}{2c_L}\right)=1+\frac{A^2+B^2}{24 c_L^2}\omega^2 \, .
\end{equation}
This raises the question of whether there can be configurations in which the plate does not emit gravitational radiation at all. In order to investigate this, we compute the total power emitted by the plate, using the standard quadrupole formula
\begin{equation}
    \dot E^{\mathrm{emit}} = \frac{1}{5}\langle\dddot M_{ij}\dddot M_{ij} - \frac{1}{3}(\dddot M_{kk})^2\rangle \, ,
\end{equation}
which, using \eqref{Mxx} and the analogue expression for $M_{yy}$, gives
\begin{equation}
    \boxed{\dot E^{\mathrm{emit}} = \frac{M^2 c_L^4 \omega^2 h^2}{15} \Big(g^2(\theta_A) + g^2(\theta_B) +g(\theta_A)g(\theta_B)\Big)} \, .
    \label{eq:emittedPower}
\end{equation}
Since the expression inside the brackets is positive-definite, the only way for the emitted power to vanish is if $g(\theta_A) = g(\theta_B) = 0$. For each $\theta$, this condition is equivalent to $\tan(\theta)=\theta+\theta^3/3$, which can be visualized in Fig.~\ref{fig:intersections}. From this plot, it becomes clear that the values of $\theta$ for which $g(\theta)=0$ become increasingly close to $\frac{\pi}{2} + n\pi$, which correspond precisely to the resonant frequencies of the plate. This raises the question of whether we can realistically probe these non-emission configurations. Let us examine a few numerical examples: the first non-zero value of $\theta$ for which $g(\theta) = 0$ is approximately $\theta_0 \approx 4.6868$, while the closest resonance happens at $\theta_{\mathrm{res}} = \frac{3}{2}\pi \approx4.7124$, yielding a difference of $\Delta \theta = \theta_{\mathrm{res}}-\theta_0 \approx 0.0256$. Comparing this to the non-emission frequency itself, we obtain
\begin{equation}
    \frac{\Delta \theta}{\theta_0} \approx \frac{0.0256}{4.6868} \approx 0.0055 \, .
\end{equation}
This ratio is independent of the material or the dimensions of the plate, and it indicates that, in order to probe the first non-emission configuration, one would need a precision of about $0.5\%$ on the frequency, which in principle is easily achievable, at least by the cryogenic aluminum resonant-mass detectors such as AURIGA~\cite{Vinante:2008zqr}, which can identify resonant frequencies to about one part in $10^6$.

These results are, of course, perturbative, and there is no a priori reason to expect them to persist at higher orders in perturbation theory. Nevertheless, it is interesting to note that, for the values of $\theta_A$ and $\theta_B$ corresponding to the black dots in Fig.~\ref{fig:intersections}, gravitational-wave emission is at least strongly suppressed. Moreover, for realistic parameter values, this suppression is, in principle, detectable. Physically, these non-emission configurations arise from a cancellation between two competing contributions to the quadrupole moment: the decrease in density resulting from the stretching of the plate, and the increase associated with the change in its physical size. At the points where these contributions exactly balance, the quadrupole moment remains constant, resulting in no gravitational radiation at linear order.

Finally, we compute the ratio between the power of the gravitational wave emitted by the plate and that of the incident wave. We need only compute the energy carried by the incident wave into the plate, which is given by
\begin{equation}
    \dot{E}^{\rm inc} = AB\frac{h^2 \omega^2}{32\pi} \, .
\end{equation}
Therefore, we have
\begin{equation}
    \frac{\dot{E}^{\rm emit}}{\dot{E}^{\rm inc}} = \frac{32\pi}{15} \frac{M^2 c_L^4}{AB} \Big(g^2(\theta_A) + g^2(\theta_B) +g(\theta_A)g(\theta_B)\Big) \, .
\end{equation}
Although this ratio is typically very small, due to the factor $\frac{M^2}{AB}$, it can become significant near the resonant frequencies.

%
%
%
\section{Conclusions}\label{section7}
In this paper, we have re-derived the model describing the interaction between a gravitational wave and an elastic body. A key feature of our formulation is that the motion of the elastic medium is governed by the standard Navier--Cauchy equations~\eqref{Navier-Cauchy}, with the gravitational wave influencing the dynamics exclusively through the boundary conditions~\eqref{boundary}, rather than through an external forcing term in the bulk equations. We then specialized to the case of a thin rectangular plate with vanishing Poisson ratio interacting with a plane gravitational wave whose propagation direction and polarization are aligned with the plate. For this configuration, we showed that the equations of motion reduce to the two decoupled one-dimensional wave equations~\eqref{wave_equations}, supplemented by the boundary conditions~\eqref{boundaryxA}--\eqref{boundaryy0}. This reduction makes it possible to obtain explicit expressions for the displacement fields and their derivatives, given by equations~\eqref{xiprime}, \eqref{xi}, and \eqref{xidot}, directly in terms of the gravitational wave signal.

We further computed the energy transferred from the gravitational wave to the elastic body in two regimes: a short burst, Eq.~\eqref{DeltaEburst}, and a long-duration harmonic signal, Eq.~\eqref{DeltaEcontinuous}. In the latter case, we employed Abel regularization to describe the physically relevant regime in which the duration of the signal is much longer than the oscillation period of the plate but remains shorter than the dissipation timescale. Finally, we determined the gravitational radiation emitted by the oscillating plate in response to a continuous harmonic wave, as described by Eqs.~\eqref{eq:GWemissionZ} and~\eqref{eq:emittedPower}. Although the plate oscillates at the same frequency as the incident wave, we found that the emitted gravitational radiation vanishes identically for certain values of the system parameters. This unexpected result can be traced to a cancellation between the contributions arising from the oscillating density and those associated with the time-dependent physical dimensions of the plate.

Several interesting directions for future work emerge from the present analysis. An important extension would be to incorporate dissipative effects, such as viscosity and internal damping, which would regularize the resonant divergences found in the idealized model and provide a more realistic description of elastic materials. Further generalizations could involve more complex constitutive relations, anisotropic or inhomogeneous media, and fully three-dimensional elastic bodies with nontrivial geometries.

A further direction for future research would be to consider more general gravitational wave backgrounds, including arbitrary angles of incidence, different polarization states, wave packets, and nonlinear gravitational waves. It would also be interesting to examine whether similar cancellation effects arise in more realistic detector models or in astrophysical elastic media, such as neutron star crusts. More broadly, the framework developed here could provide a basis for constructing analytically tractable relativistic models of gravitational wave detectors and for further elucidating the connections between elasticity theory and gravitational wave physics.

%
\section*{Acknowledgements}
This work was partially funded by FCT/Portugal and the Recovery and Resilience Plan (PRR) through projects UID/04459/2025 (DOI: 10.54499/UID/04459/2025) and UID/PRR/04459/2025 (DOI:10.54499/UID/PRR/04459/2025), by the European Union H2020 ERC Advanced Grant ``Black holes: gravitational engines of discovery'', GA no.~Gravitas–101052587, and also by the H2020-MSCA-2022-SE project EinsteinWaves, GA no.~101131233.
%
%

\end{document}